# Thin-wall Single-crystal Gold Nanoelectrodes towards Advanced Chemical Probing and Imaging

Authors: Milad Sabzehparvar[1], Fatemeh Kiani[1], Germán García Martínez[1], Omer Can Karaman[1], Victor Boureau[2], Lucie Navratilova[2], Giulia Tagliabue[1*]

[1] Laboratory of Nanoscience for Energy Technologies (LNET), STI, École Polytechnique Fédérale de Lausanne, 1015 Lausanne, Switzerland

[2] Interdisciplinary Center for Electron Microscopy (CIME), École Polytechnique Fédérale de Lausanne, 1015 Lausanne, Switzerland

*E-mail: giulia.tagliabue@epfl.ch

**Abstract**

Thin-wall metal ultramicro- and nanoelectrodes (UMEs/NEs), especially gold NEs, are indispensable for high-resolution electrochemical microscopy, biosensing, and fundamental research. However, their damage susceptibility and the lack of scalable fabrication methods hinder broader adoption. We present a versatile wet-chemical approach for high-throughput fabrication of thin-wall Au NEs/UMEs and multifunctional NEs with ~80% reproducibility. This method is based on a unique template-assisted 1D growth of single-crystalline Au in borosilicate nanopipettes followed by electrochemical contacting with tungsten microwires, and focused ion beam milling, ensuring precise control over NEs dimensions. Adaptable to various metals and integrable in multifunctional probes, the method facilitates batch production of high-quality NEs with standardized electrical connections. Structural and electrochemical characterization reveals a twinned single-crystalline Au core, a seamless Au/glass interface, and highly stable electrochemical performance. Notably, smaller electrodes exhibit higher current densities, enhancing chemical detection sensitivity. Specifically, we demonstrate outstanding spatial (< 200 nm) and current (< 1 pA) resolutions, low limit of detection (~11.0 μM) and high stability (7 h) in scanning photoelectrochemical microscopy (photo-SECM), by detecting photo-oxidation reaction on atomically smooth Au micro-flakes. We also demonstrate growth in double-barrel pipettes for SECM/SICM probes as well as Pt NEs. Overall, this scalable method addresses longstanding challenges in NEs, paving the way for advanced electrochemical and spectro-electrochemical microscopy, including SERS/TERS integration. With single-crystalline surfaces, these electrodes open new frontiers in catalysis, interfacial electrochemistry, biosensing, and molecular-scale investigations.

**Keywords:** High-throughput Fabrication, Metal Nanoelectrodes, Electrochemical Microscopy, Near-field Microscopy, Electrochemical Sensor, Biosensor, Crystal Growth

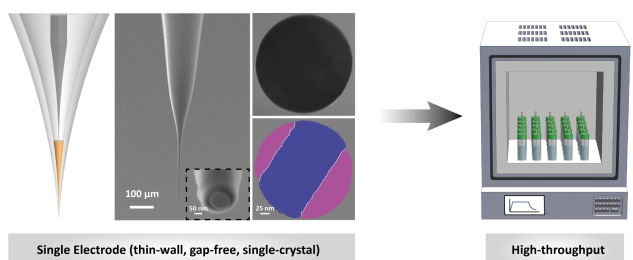

**Introduction**
Nanoelectrodes (NEs), with radii below ~100 nm, serve as powerful probes for investigating electrochemical processes at the nanoscale[1,2]. This is largely due to their size being much smaller than the diffusion layer thickness, resulting in extremely high mass transport rates, low ohmic losses, small RC time constants, and fast steady-state responses.[1,3] These unique advantages enable investigation of numerous phenomena and processes that cannot be studied at macroscale electrodes, including electrochemistry of individual molecules[4–6] and nanoparticles[7–10], formation and growth of transient metal nuclei[11,12], nanobubbles[13–15], short-lifetime intermediates[16,17], rapid heterogeneous electron transfer kinetics[18–20], and non-invasive electrochemistry inside living cells[21]. Among all metals, gold NEs find more extensive applications in bio-analysis[22–24] and catalysis[25] research due to their inertness, high surface reactivity and higher affinity for biomolecules and chemical bonds (**Figure 1a**).

Laser pulling of glass-sealed metal microwires is a common technique for fabricating disc-shape metal NEs.[26,27] However, despite the huge improvements in the past decades for achieving platinum ultramicro-/nanoelectrodes (UMEs/NEs)[26,28], challenges still exist to reproducibly and massively produce high-quality, "leak-free", gold NEs. This is primarily due to the large difference between the melting point of gold (1064 °C) and quartz (1710 °C), resulting in low success rate for pulling continuous gold NEs and a poor seal at the gold/glass interface. This leads to an unstable electrochemical performance[18] and necessitates transmission electron microscopy (TEM) for controlling the fabrication[29], severely limiting the throughput of the process. In addition, laser-pulled electrodes, while exhibiting some degree of grain alignment, are usually polycrystalline with randomly oriented crystallites and grain boundaries[30,31]. Most importantly, the resulting NEs typically feature a thick glass insulation sheath, with a glass-to-conductive core radius ratio (known as the RG value) substantially exceeding 10, resulting in overall dimensions within the micrometer scale.[27]

Nanoelectrodes with small RGs, e.g. 1.2-2, have emerged as critical tools for probing neurotransmitters, and achieving ultrahigh spatial resolution in chemical[32,33] and spectroscopic (near-field) imaging[34–37]. This originates from their small footprint, minimizing the mechanical damage to a cell in neurobiology and allowing extremely small tip-to-substrate distances in scanning electrochemical microscopy (SECM), critical for enhanced detection sensitivity[32]. Despite huge efforts in advancing the fabrication of such NEs, for example by using a thin polymer coating on metal micro-/nanowires[38], or by reducing the thickness of the glass sheath of laser-pulled NEs by chemical etching[39], mechanical polishing[40], or local heating[33,41], these processes remain either very time-consuming and challenging or frequently result in a defective insulating sheath having parasitic current leakage.

Glass nanopipettes, instead, are thin-wall nano-pore structures that are nowadays easily fabricated with sub-50 nm orifice sizes at uniquely high yield and repeatability[42–44]. Recently, room-temperature template-assisted chemical and electrochemical deposition of metals in glass nanopipettes has been found as a promising approach for controlled fabrication of low-RG-value metal NEs[45–48]. However, crystal growth is a self-limited process in all these methods, resulting in short-length deposits and limiting the performance only to a wireless bipolar electrochemical contacting approach[49,50]. Moreover, the yield of these serial fabrication processes remains low, and poor sealing at metal/glass interface, especially for larger-size NEs[48], remains an issue, leading to a high background ionic current in addition to the Faradaic current detection. Thus, it is necessary to develop a high-throughput processing method for deposition of very long length Au crystals and realization of high-quality physically-contacted thin-wall Au NEs.

In this work, we present a robust and versatile wet-chemical method for the scalable fabrication of thin-wall metal NEs/UMEs with a perfect disc-shape geometry and precise control over size and RG values (**Figure 1b-d**). Our approach leverages template-assisted 1D growth of single-crystalline gold within borosilicate nanopipettes using a polyol-based process, followed by standard physical contacting using a tungsten micro-wire and focused ion beam milling. The polyol process is pivotal to this method, enabling the continued and selective growth of large-size metal deposits within the nanopipette. The resulting electrodes uniquely exhibit a twinned single-crystalline structure, a seamless Au/glass interface, and stable electrochemical performance both in electrochemical environments and after long-term storage in air. This technique enables the production of electrodes with radii ranging from nanoscale (~50 nm) to microscale (~500 nm) and higher, along with multifunctional designs (e.g. dual, double-barrel, and recessed L-shape NEs), extending its applicability to other technologically important metals like Pt, Ag, Cu, Pd, and even Bi (**Figure 1b**). Importantly, smaller electrodes demonstrate higher current densities, highlighting their enhanced sensitivity for chemical detection at exceptionally high spatial resolutions. We demonstrate the potential of these Au NEs in advanced electrochemical applications, including photo-SECM, where they easily detect a weak photo-oxidation reaction over atomically smooth Au microflakes at subwavelength spatial resolution (< 200 nm), record-low current step (~ 1 pA) detection, and outstanding detection limit (~11.0 μM), and with a high stability (**Figure 1a**). Overall, this high-throughput method addresses key challenges in nanoelectrode fabrication and electrochemical imaging, and opens completely new avenues for high-performance nanoelectrochemical technologies, including coupling to near-field spectroscopy methods[51,52], such as SERS/TERS, towards advanced fundamental research in (photo)electrochemistry, such as double-layer structure and biomolecule interactions[15,53].

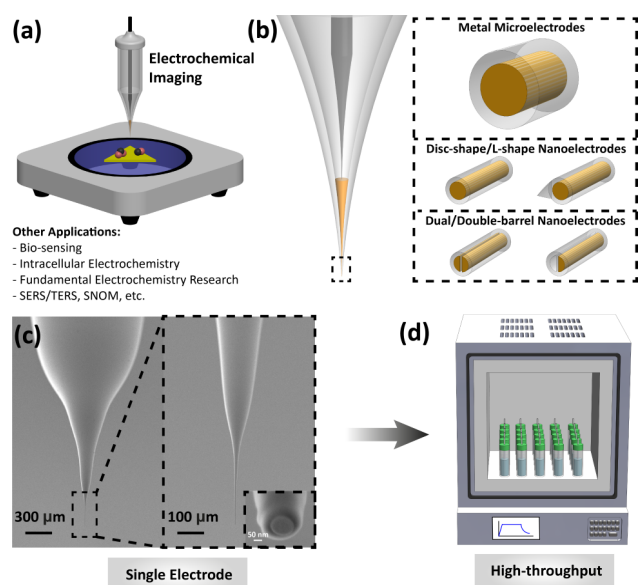

**Figure 1.** Thin-wall metal nanoelectrodes. (a) Schematic illustration of the implementation of the electrodes for electrochemical microscopy technologies (e.g. SECM, EC-STM, and EC-AFM), and other potential applications. (b) Schematic illustration demonstrating flexibility of the wet-chemical approach in obtaining various physically-contacted thin-wall electrochemical probes having different material, geometry, and sizes. Using salt of different metals can enable fabrication of different metal NEs. Complete chemical growth of gold metal inside single-barrel and asymmetric double-barrel nanopipettes enables high-throughput fabrication of disc-shape gold micro-/nanoelectrodes and multi-functional nanoelectrodes. A partial cross-sectional FIB cutting enables fabrication of recessed L-shape metal nanoelectrodes. (c) SEM images of a typical thin-wall gold nanoelectrode with a disc-shape geometry. (d) Schematic illustration demonstrating the mass production of metal nanoelectrodes using a wet-chemical growth approach inside a muffle furnace. Multiple glass vials are processed simultaneously within the furnace, showcasing the scalability of the fabrication method for large-scale production of nanoelectrodes.

**Results and Discussion**

A thin-wall (i.e. low-RG-value) NE is defined as an electrode with an RG value smaller than 10, preferably less than ~2, which corresponds to a wall thickness equal to the core radius and an optimum back-diffusion of chemical species. Fabrication of thin-wall Au NEs was performed by employing a polyol-based wet chemical growth process inside glass nanopipette templates, recently reported for 2D growth of single-crystalline Au micro-flakes on glass substrates[54]. The process involves three main stages as illustrated in **Figure 2a**. Briefly, borosilicate glass nanopipettes are fabricated by a laser pulling process (**Figure 2a.i**) and used as templates for selective growth of long-length gold deposits at the very end of the nanopipette (**Figure 2a.ii** and **Figure 2b-c**) followed by its physical contacting to an electrochemically sharpened tungsten micro-wire under an optical microscope (**Figure 2a.iii**). A focused-ion beam (FIB) milling is finally performed to precisely cut the protruded Au deposit and realize perfect disc-shape Au NEs. We note, however, that our growth method can result in NEs with a nanosized conical protrusion, excluding the need for the FIB step for many applications, e.g. biosensing[23] and spectroscopic[37,55] imaging, that do not need a disc geometry.

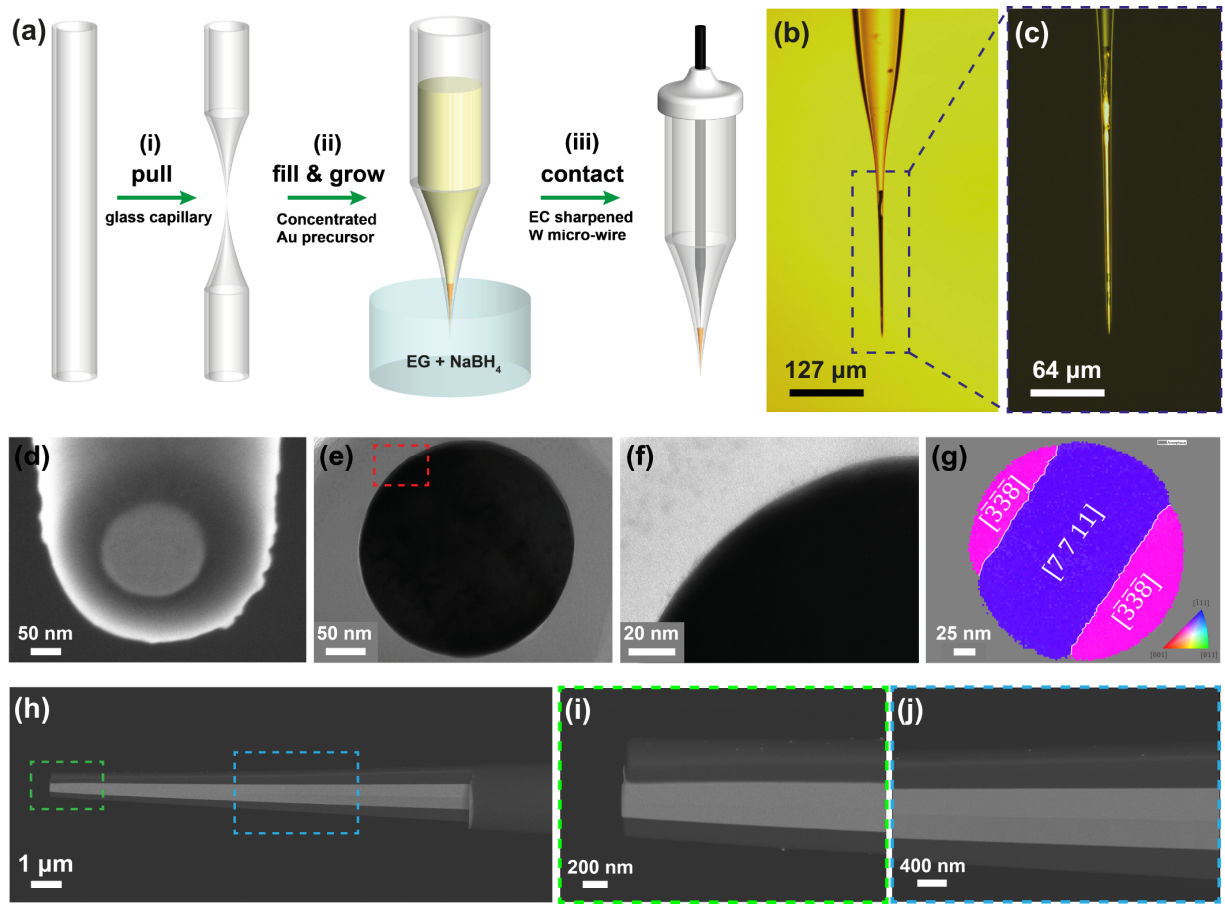

**Figure 2.** Fabrication process and quality analysis for thin-wall single-crystalline gold nanoelectrodes. (a) Schematic illustration of wet chemical growth of thin-wall gold nanoelectrodes. (i) Pulling of glass capillaries into sub-20 nm glass nanopipettes. (ii) Filling the nanopipettes with a growth solution containing $AuCl_4^-$ ions in ethylene glycol and insertion of the filled nanopipette into a bulk media containing ethylene glycol and sodium borohydride. Huge gold crystal grows inside the nanopipette via a polyol-assisted process. (iii) Physical contacting by connecting an electrochemically sharpened tungsten micro-wire to the large-size gold deposit. Transmission (b) and reflection-mode (c) optical micrographs of an as-grown >200 μm long, gold deposit inside the glass nanopipette. (d) Tilted-view SEM image of a FIB-cut gold nanoelectrode at 54° angle. (e) Top-view BF-TEM image of a sliced lamella of gold nanoelectrode. (f) High-magnification BF-TEM image of the gold-glass interface marked with a red dashed rectangle in (e). (g) Orientation map of the Au crystal with respect to the axis of the nanoelectrode shown in (e), with {111} Σ3 twins overlayed in white. The inset shows the corresponding color legend. (h-j) SEM images of an axially FIB cut Au NE measured by an in-lens backscattered electron detector. They show the perfect sealing and single-crystallinity of the gold deposit along the glass capillary. The Au NE was coated with a 10 nm carbon layer for decreasing the charging and drifting problem during the FIB cutting process.

Ideal-shape glass nanopipettes are fabricated by a unique combination of laser-shrinking and laser pulling of borosilicate glass capillaries having large inner diameters that results in short-taper (~300 μm) and ultrafine, sub-20 nm opening nanopipettes with high reproducibility (see **Methods**). Importantly, the laser-assisted pre-shrinking process enables fabrication of exceptionally small borosilicate nanopipettes by locally increasing the glass wall thickness, overcoming the intrinsic softness problem of this material. Depending on the duration of the shrinking time and after a delayed hard pulling process, the orifice size, RG value, and taper length of the nanopipette can be all finely tuned with high repeatability. It is to note that glass nanopipettes are commercially available with different RG values, but our approach enables fabrication of ultra-fine and short-taper nanopipettes out of thin-wall capillaries having a large inner diameter, as well as RG values (e.g. ~6) that are not available commercially, which facilitates all subsequent steps of the process. This direct method improves the control and reliability of the ultrafine nanopipette characteristics compared to previously reported secondary post-processing approaches, such as electron-beam shrinking[56] or atomic layer deposition of insulating coatings[57]. Critically, an exceptional (~90%) reproducibility was achieved for fabrication of ideal-shape borosilicate nanopipettes, ensuring the overall reliability of the nanoelectrode fabrication method.

For fabrication of Au NEs (see **Methods**), the glass nanopipettes are back filled with a growth solution containing $AuCl_4^-$ ions in ethylene glycol (EG) using a home-built glass injector, and then immediately inserted vertically in glass vials containing EG and sodium borohydride ($NaBH_4$). Next, the vials are promptly transferred to a furnace heated at 110 °C for accelerating the gold deposit growth via a polyol synthesis process. During the heating process, the aldehyde groups (−CHO) of the heated EG continuously reduce the $AuCl_4^-$ ions into gold nanocrystals[54] whose directional growth is then accelerated by the strongly reducing $BH_4^-$ ions, uniquely providing a continued and selective growth even after the complete blockage of the orifice with metal deposit. We performed several concentration studies revealing that (i) gold nucleation is initiated by a confinement effect close to the orifice rather than randomly across the nanopipette body, (ii) diffusion of both $NaBH_4$ and $AuCl_4^-$ is critical for the growth process, the latter determining the morphology of the growing crystal, and (iii) a time-dependent concentration (increasing) of $AuCl_4^-$ enables selective growth of long gold deposits while minimizing unwanted nanoparticle on the outer surface of the glass capillary. To achieve the latter, the nanopipette tip is thus first dipped in pure EG while a high $AuCl_4^-$ solution is backfilled. We also observed that optimized growth conditions are necessary to avoid a severe chemical attack to the thin glass close to the nanopipette apex at both high $AuCl_4^-$ ions and $NaBH_4$ concentrations, an effect that is strongly enhanced by increasing the growth temperature and time. The final FIB cutting step allows exposure of a disc-shaped gold electrode with perfect sealing (**Figure 2d**) and if needed, removal of any damaged glass section. Overall, we found that 0.125 M $AuCl_4^-$ in EG growth solution, 0.5 ml of 200 mM $NaBH_4$ in ethanol in 2 ml EG bulk media, 110 °C growth temperature and 24 hours growth time are optimized conditions for obtaining small-size Au NEs. When using ~20 nm radius glass nanopipettes as starting point, this process results in ~ 80% success rate for achieving around 300 µm continuous gold deposits completely filling the conical part of the nanopipette with the least amount of unwanted growth residues in the pipette stem and on the exterior surfaces (see **Figure 2b,c**). After FIB cutting, a sub-100 nm-radius Au NE can be reliably obtained.

The ~300 µm typical gold deposit length obtained with the described method is about 130, 10, and 15 times higher than the ones previous obtained by bipolar electrochemistry[47], electrochemical deposition[46] and interfacial reaction[45] methods, respectively. Together with the optimized nanopipette short-taper, it uniquely enables standard physical contacting of the NEs using metal wires. This approach is indeed critical to overcome unavoidable limitations of bipolar electrochemical contacting such as poor long-term stability and current-concentration nonlinearity[49,50,58]. We also optimized a two-step, static electrochemical etching approach for facile fabrication of long-length (700 µm taper length) tapered tungsten micro-wires (5 µm width, **Methods**) with controllable profiles and very high (>90%) reproducibility. This W wire profile was found ideal in size and mechanical strength for successfully passing through the narrow glass body of the Au NEs and achieving soft electrical contacting to the long-length Au deposit in a spring-loaded condition. The whole contacting process was monitored under an optical microscope and was successful with a very high repeatability (> 90%). It is to note that the W etching step can be even skipped as our chemical growth approach can exceptionally grow giant gold crystals for electrical connection using commercially available 25 µm-diameter (or smaller) W micro-wires.

Quality of the fabricated electrodes was assessed by electron microscopy and via an electrochemical analysis approach, to resolve details of the physical and chemical properties, respectively. **Figure 2d** shows an SEM image of a typical Au NE after physical contacting and FIB cutting processes. The Au disc is highly circular with a geometrical radius of 85 nm, and has a seamless interface with a pinhole-free and thin glass sheath. Axial FIB cutting on a different Au NE showed the gold/glass interface remains perfect, i.e. seamless, even up to tens of micrometers from the very end (**Figure 2h-j**). This indicates a complete chemical growth of Au, filling its template shape, and promises a large flexibility for the fabrication of Au NEs having different sizes by this approach (see **Figure 4a-e**), from microelectrodes to nanoelectrodes. Bright-field (BF) TEM imaging of a thin FIB cross-section of the electrode shown in **Figure 2d** further confirmed the seamless gold/glass interface with no gaps (see **Figure 2e,f**), indicating a complete chemical growth for the Au deposit. A precession-assisted TEM experiment for crystal orientation mapping[59] (see **Figure 2g**) showed the existence of characteristic twins in gold, precisely Σ3 twin boundaries, characterized by a 60° rotation around <111> crystal axis [60]. Moreover, the twinning planes are parallel to {111} planes for crystal grains at both sides, revealing coherent twins, precisely {111} Σ3 grain boundaries. Crystal domains are oriented [7,7,11] and [-3,-3,-8] along the NE axis, respectively for the bigger and the two smaller domains. The orientation mapping on a bigger sized NE prepared by FIB cross-sectioning upper in the deposit showed the same {111} Σ3 twinned boundary structure with increased number of crystal grains, showing [7 5 11] and [-6,-4,-11] orientations along the NE axis, respectively for the bigger and for the smaller domains. These orientations reveal a preferred close-to-<111> orientation, as highlighted by the plot of the inverse pole figures along the NE axis where the intensities are distributed closer to the [111] pole. Precisely, the misorientation of the <111> crystal axis of the larger grain with respect to the NE axis is 18.0° for the bigger NE, but only 12.7° for the smaller NE. The authors anticipate that even smaller nanoelectrodes (<100 nm radius), if successfully observed by TEM via cross-sectional preparation, would exhibit crystallographic orientations even closer to <111>. Some clues to understand this phenomenon can be related to a slight change in the crystal growth trajectory during the growth process inside the conical nanopipette channel observed at micrometer scale, or position changing of the twins inside the Au disc observed at nanometer scale with serial FIB cross-sectioning as a result of a misalignment between the <111> crystal axis and the pipette axis because twins were observed to lay in the {111} planes. Further studies are needed to investigate the underlying growth mechanism as well as its role in the formation of the record-large metal deposit achieved through our polyol-based method.

To test the electrical contact of the NEs and electrochemically examine their geometrical size we performed cyclic voltammetry experiments in a redox couple solution[61] (see **Methods**). **Figure 3a** shows the voltammetric response of the Au NE in **Figure 2d** at two different scan rates in an electrolyte solution containing 2 mM $Fe(CN)_6^{4-}$ in 0.5 M $Na_2SO_4$. It can be seen that the electrode exhibits a well-defined sigmoidal CV with a plateau diffusion limiting current and little hysteresis at low scan rates, i.e. 10 mV.s$^{-1}$. Importantly, it remains in this steady-state condition even at scan rates as high as 0.5 V.s$^{-1}$. This indicates the good electrical contact and high surface reactivity of the NE, and confirms the absence of a solution-filled gap at the gold/glass interface.[18,62] Considering a disc-shape NE with thin glass walls, the radius of the electrode can be calculated from the diffusion-limited steady-state current:

$i_{T,\infty} = gnFDC^*r$ (eq. 1)

where n is the number of electrons transferred, F is Faraday's constant, r is the core radius, D and $C^*$ are the diffusivity and bulk concentration of $Fe(CN)_6^{4-}$ molecules, respectively. The geometrical factor g is dependent on the glass wall thickness, as follows,

$g = 4.00 + 4B(RG - C)^D$ (eq. 2)

Using the literature-established values of B = 0.1380, C = 0.6723, D = −0.8686,[63,64] and an RG value of 2 for the electrode, the measured 40 pA limiting current corresponds to an estimated electrode radius of 85.7 nm that closely matches the geometrical radius estimated by SEM (~83 nm).

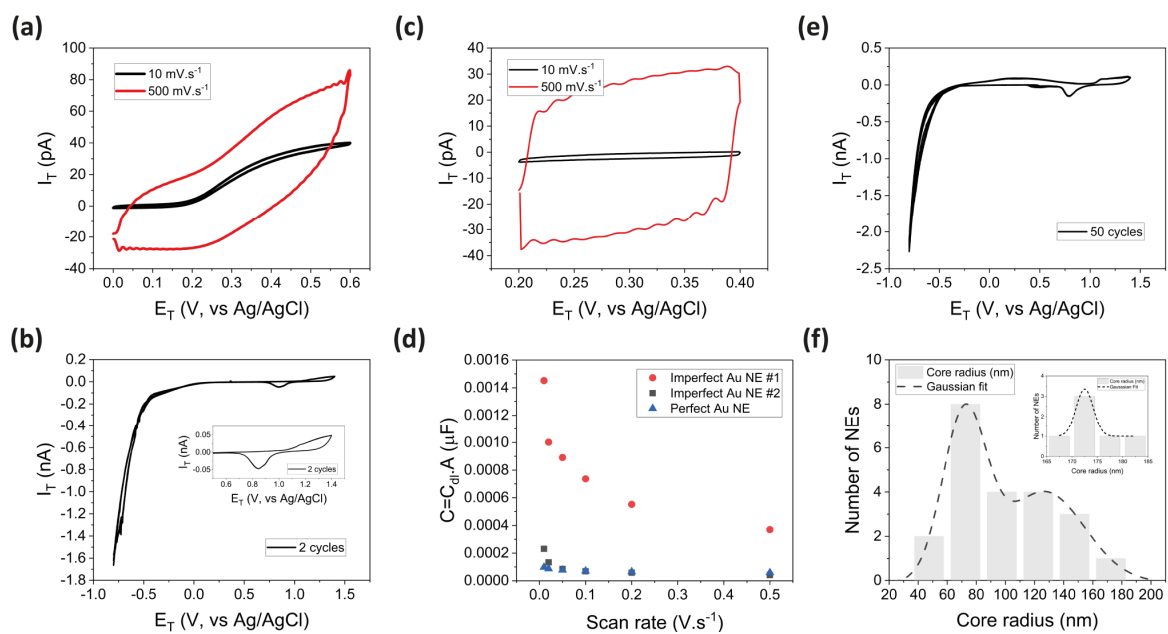

**Figure 3.** Electrochemical characterization, electrode stability and process reproducibility. (a) Cyclic voltammograms of the nanoelectrode in Figure 2d in an electrolyte solution containing 2 mM $Fe(CN)_6^{4-}$ redox molecules in 0.25 M $Na_2SO_{4, (aq.)}$ at low and high scan speeds. (b) Two-cycles cyclic voltammograms of the nanoelectrode, from HER to gold redox potential windows, at 10 mV.s$^{-1}$ in a 50 mM $H_2SO_4$ electrolyte. The inset shows perfect overlap of the gold redox peaks over two consecutive cycles of CVs up to the HER triggering regime, i.e. -0.8 V vs Ag/AgCl. (c) Cyclic voltammograms of the nanoelectrode at 10 and 500 mV.s$^{-1}$ scan speeds within the double-layer potential window. (d) Plot of total capacitance with scan rate for the perfect Au NE in Figure 2d and imperfect NEs. (e) Stability of a ~80 nm-radius Au NE over 50 cycles (~70 min) at 50 mV.s$^{-1}$ in a 50 mM $H_2SO_4$ electrolyte. (f) Statistical analysis results for disc-shape nanoelectrodes fabricated in the same batch (total of 26 empty nanopipettes) by the wet chemical approach. The inset shows analysis result for disc-shape Au NEs obtained by controlled FIB cutting for a specific electrode radius, e.g. ~170 nm.

While outer-sphere voltammetry is not very sensitive to surface composition and nanoscale imperfections, blank voltammetry in $H_2SO_4$ solution has been identified as the most reliable method for evaluating the surface chemistry, the electrochemical surface area and the roughness quality factor (RF value) of the electrodes.[61,65] **Figure 3b** shows two consecutive CV cycles of the same NE in a 50 mM $H_2SO_4$ solution at 10 mV.s$^{-1}$ scan rate in a potential window covering both the hydrogen evolution reaction (HER) and gold redox processes. The presence of clear gold oxidation and reduction peaks at 1.4 V and 0.84 V vs Ag/AgCl is indicative of a clean gold surface.[18,61] From the consumed charge under the gold reduction peak (390 μC.cm$^{-2}$), we estimated a RF value of ~7, in line with those of typical NEs (between 1.2 and 12), due to likely deviations from a 1:1 stoichiometric monolayer formation of gold oxide/hydroxyl and its reduction at nanoscale.[66–68] The sealing quality (gold/glass interface quality) of the NEs, can be further confirmed by two features of the CV scans: (i) the highly retraceable CV curves in the HER and gold redox potential regimes (**Figure 3b**), as any defect would lead to unreproducible measurements, and (ii) the flat profile of the CV baseline within the double layer regime up to scan rates as high as 0.5 V.s$^{-1}$ (**Figure 3c**), as defect would result in tilted scans due to resistive contributions.[65,69] Additionally, the negligible (<2 times) dependency of the calculated double-layer capacitance ($C_{dl}$) on the scan rate (**Figure 3d**), especially at scan rates smaller than 100 mV.s$^{-1}$, further supports the gap-free nature of the gold/glass electrochemical interface, as defects could result in orders of magnitude increase in the $C_{dl}$ value due to the variation of the iR drop.[69] Indeed, experiments on selected defective NEs having different gap sizes confirm for the first time the sensitivity of these assessment techniques to sealing quality (**Figures 3d**).

Stability of the electrodes was also evaluated by performing 50 cycles of CVs within the HER and gold redox regimes, corresponding to ~75 min of electrochemical experiment in 50 mM $H_2SO_4$ solution. As can be seen from **Figure 3e**, there is no obvious change in the CV response of the nanoelectrode, further confirming the sealing quality assessment with electrochemical analysis tests. Nonetheless, SEM imagining before and after the cycling treatment shows a slight recession for the NE due to electrochemical etching of gold during its redox process. While this is an inevitable damaging mechanism for small NEs[70], in our tips it interestingly happens in a smooth, layer-by-layer manner due to the single-crystalline nature of the grains, and the low-energy coherent nature of the {111} Σ3 twin boundaries in our Au NEs[60]. Thus, the twin boundaries are still evident on the tested electrodes and, most importantly, the gold/glass interface does not deteriorate. This allows prolonged operation and uniquely enables the possibility of easy re-using the electrodes after a recovery FIB cutting process. Notably, the electrodes exhibited nearly identical electrochemical performance even after long-term storage (e.g., over three months) in air, demonstrating exceptional stability in electrical connection and surface reactivity.

Overall, our comprehensive physical and electrochemical characterizations indicate the high quality of the fabricated NEs, ensuring a stable electrochemical performance for advanced scientific research. This represents a significant breakthrough in the fabrication of perfect thin-wall Au NEs, overcoming longstanding challenges. Unlike secondary electrochemical deposition methods using recessed NEs, which often result in porous Au structures with inconsistent and unpredictable electrochemical behaviour[23,71], our approach delivers exceptional performance in attaining these highly sought-after solid and gap-free thin-wall disc-shape electrodes.

To quantify the reproducibility, reliability and scalability of our fabrication protocol, we measured the size distribution of one batch of fabricated Au NEs from their SEM images (**Figure 3f**). In this instance we pulled 13 glass capillaries into 26 identical nanopipettes having a ~20 nm-radius orifice size (> 90% yield). After the chemical growth, physical contacting and FIB cutting processes, we obtained 22 Au NEs with large-size Au deposit and perfect disc-shape geometry (> 80% reproducibility). The discarded NEs were either having a relatively shorter Au deposit or a slight gap at the gold/glass interface. Remarkably, the size distribution is very narrow, spanning from ~55 nm to ~170 nm. Importantly, the as-grown size distribution can be further narrowed by controlling the cutting size during the FIB cutting step towards a specific electrode size, e.g. 175±10 nm (see the inset in **Figure 3f**). This shows that the reported approach offers high throughput and controllable fabrication of electrodes having a specific size that is highly critical in many applications.

An interesting possibility opened by our approach, is the direct comparison of the electrochemical performance of Au NE with different radii. In fact, the size of the Au NE depends on the nanopipette template and the single crystalline nature of the Au offers excellent control onto the tip material. We thus repeated our detailed quality assessment on disc-shape Au NEs having sizes ranging from ~45 nm to ~550 nm radii. **Figure 4a-e** show SEM images of the studied electrodes. As can be seen from **Figure 4f**, all the electrodes show ideal sigmoidal shape voltammograms with low hysteresis even for the smallest NE, whose electrochemically calculated radius is similar to the geometrical values measured from SEM images. Normalization of the CVs to the limiting current values of the oxidation waves revealed no obvious size-dependency for the half-wave potential, $E_{1/2}$, and thus surface reactivity for $Fe(CN)_6^{4-}$ oxidation on different size electrodes. This is in agreement with the surface insensitivity of this fast reaction based on outer-sphere charge transport mechanisms[65,72].

**Figure 4g** shows the CV response of the studied electrodes in 50 mM $H_2SO_4$ solution. The electrodes show well-defined gold redox peaks with a lower current value for smaller electrodes. Interestingly, a significant positive potential shift of the gold reduction peak is observed by decreasing the electrode radius from ~550 nm to 75 nm, followed by a negative shift for the 45 nm-radius NE. While the observed negative shift for the ~45 nm-radius NE is in agreement with the previous reports on Au NEs[73] and Ag NPs[74] and is attributed to a change in the standard potential of Au and Au oxide, we could not clarify the observed positive shifts for larger electrodes and it will be the subject of future research. Testing the electrodes in the HER regime (**Figure 4h**) interestingly showed that the HER onset potential also decreases with the electrode size, resulting in an enhancement of the current density (**Figure 4h**). A two-slope HER regime was observed for sub-100 nm NEs, i.e. the 75 nm and 45 nm samples, corresponding to an enhanced H adsorption before triggering of HER[75]. All studied electrodes showed a stable voltammetry response during cycling up to the HER potential regime, a flat CV response at high scan rates, and a negligible (< 2 times) dependency for $C_{dl}$ to the scan rate, indicating a high quality for the gold/glass interface.[69] Notably, a higher $C_{dl}$ value was calculated for smaller size electrodes,[69] in line with our observed stronger H adsorption/desorption processes that contribute in the double layer structure formation, and theoretically calculated electrode curvature/edge effects[76]. We propose that, in addition to size-related effects, all the observed trends in redox potential, HER activity, and $C_{dl}$ values may share a common origin: the varying crystallographic orientations of the electrodes.[77,78] Smaller electrodes, with orientations closer to [111], likely exhibit a higher density of atomic interactions, which could account for the observed behaviour. Stability analysis on larger-sized electrodes revealed improved stability with less recession depth.

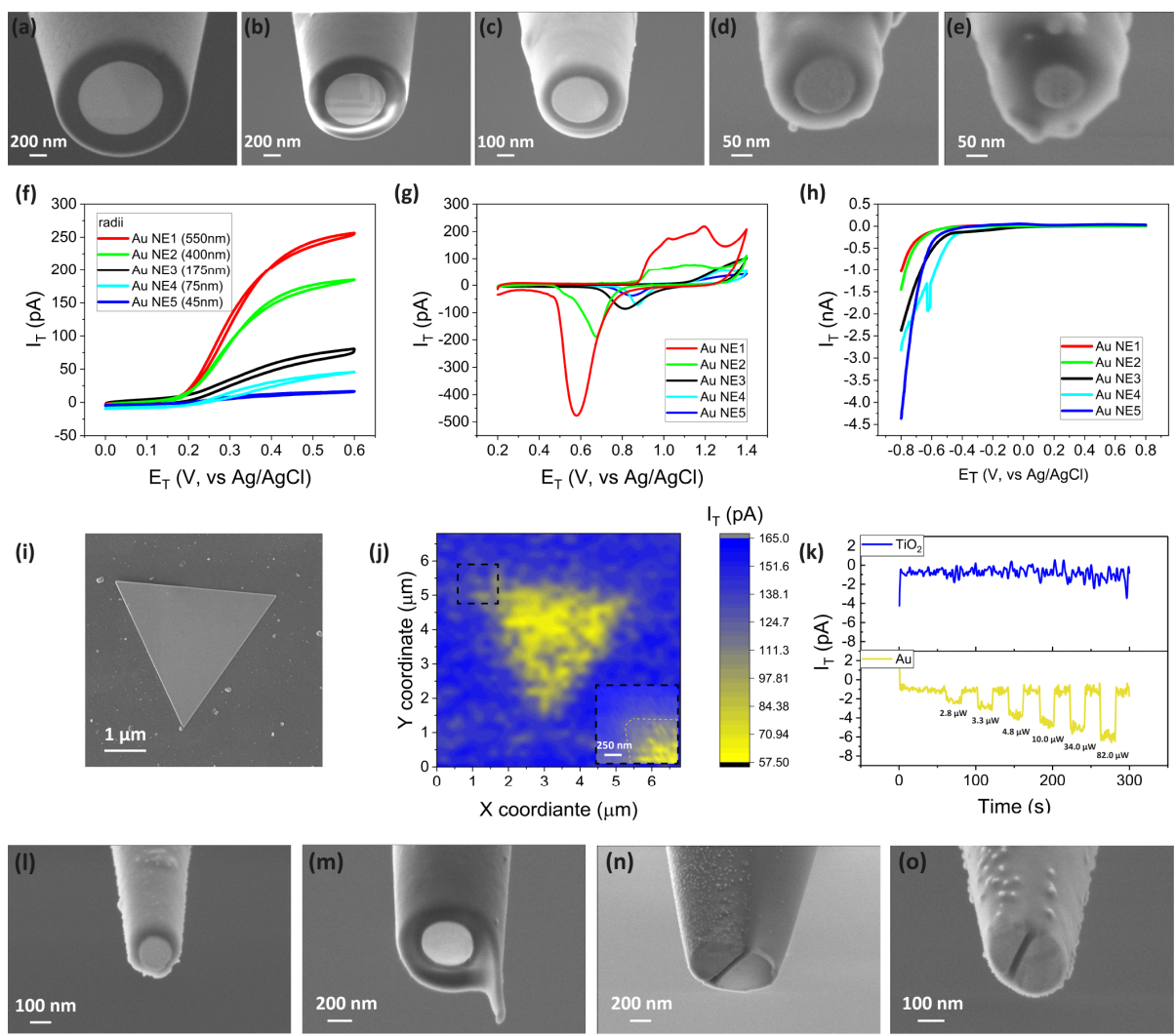

**Figure 4.** Size effect, photoelectrochemical imaging and electrode versatility. (a-e) Tilted-view SEM images of disc-shape Au NEs having different core sizes. (f) Cyclic voltammograms of disc-shape gold nanoelectrodes having different radii in an electrolyte solution containing 2 mM $Fe(CN)_6^{4-}$ in 0.25 M $Na_2SO_4$. Cyclic voltammograms of the gold nanoelectrodes in 50 mM $H_2SO_4$ within the gold redox (g) and HER (h) regimes. (i) Top-view SEM image of a typical ~20 nm-thick triangular single-crystalline Au micro-flake on $TiO_2$/ ITO substrate, analogous to the one imaged in panel (j); (j) Constant-height photo-SECM image of photo-oxidation of $Fe(CN)_6^{4-}$ molecules on a gold micro-flake surface. The image was obtained with a 200 nm-radius Au NE biased at a 0.4 V vs Ag/AgCl in a 4 mM $Fe(CN)_6^{4-}$ in 0.25 M $Na_2SO_4$ solution, a 516 nm laser excitation (10 µW), and at 200 nm scan line spacings. The inset shows a higher resolution (100 nm) photo-SECM image of the corner of the flake marked with a red dashed rectangle. No post-processing is applied to the spatial data. (k) Time traces of tip current obtained from the Au NE upon local illumination of the $TiO_2$ substrate and Au micro-flake with an excitation wavelength of 516 ± 10 nm at different light intensities (2.8-82 µW). The tip was biased at 0 V vs Ag/AgCl reference electrode in substrate generation-tip collection mode of operation. (l-m) Tilted-view SEM images of (l) a platinum nanoelectrode, (m) a recessed L-shape nanoelectrode, (n) a double-barrel gold nanoelectrode (left barrel: Au, right barrel: empty), and (m) a dual gold nanoelectrode.

To further validate the functionality of our Au NE in real SECM experiments, we performed imaging of a photo-oxidation reaction on ~20 nm-thick single-crystalline Au micro-flakes (Au MFs) on a TiO$_2$ substrate[54] using a 215 nm-radius Au NE tip (**Figure 4i** for a representative MF). Specifically, we immerse the flakes in a solution of 4 mM Fe(CN)$_6^{4-}$ in 0.25 M Na$_2$SO$_4$. We electrochemically position the tip approximately 315nm from the TiO$_2$ substrate, where strong feedback happens between the tip and substrate. We photoexcite the unbiased sample using a focused 516 nm laser (10 µW, backside illumination), collinear with the Au NE tip position. Under illumination, the photo-generated hot electrons in Au MF inject into the TiO$_2$ conduction band while photo-generated holes participate in the photo-generation of Fe(CN)$_6^{3-}$ oxidant molecules[79]. By applying an oxidative potential of 0.4 V vs Ag/AgCl to the Au NE, we can exploit a competition mode of operation during irradiation, with the same oxidation reaction happening at the tip electrode and substrate surface. This results in a decrease in Faradaic tip current ($I_T$) on areas with higher photoactivity. As we scan the sample, we can thus build an image of the sample activity with a pixel size of 200 × 200 nm² (**Figure 4j**). We observed a fairly uniform photoactivity across the flake surface, as previously reported in dark conditions.[80] Gold micro-flakes can inherently feature truncated corners, with their size depending on the growth conditions.[54] Our higher resolution photo-SECM imaging, performed at a 100 × 100 nm² pixel size, can resolve small features such as the sub-200 nm truncated nature of the corner of the studied flake. Notably, the achieved spatial resolution of less than 200 nm surpasses the state-of-the-art for photo-SECM imaging (approximately from a few microns down to 200 nm), which is inherently limited by the illumination conditions[81–83]. Additionally, in a constant-height imaging mode we experienced no crash of the tip and an extremely stable performance for over 7 hours of electrochemical imaging at both oxidative and reductive potentials. The same Au NE could also be reused at least 5 more times after FIB cutting/imaging cycles without observing any performance degradation, albeit with a slight increase in tip size. These outstanding capabilities all originate from the thin glass wall of the electrode and the perfect seal at the glass/gold interface. It is to note that higher spatial resolutions are easily possible by using smaller size Au NEs and smaller step sizes. Moreover, our local photo-SECM measurements at varying laser powers on the TiO$_2$ and Au MF regions, using a 0 V vs Ag/AgCl tip potential (**Figure 4k**), showed the ideal operation for the Au NEs in substrate generation-tip collection mode. These measurements also highlighted the sensitivity of the tip electrode in chemical sensing, with higher $I_T$ values observed at increased laser powers on the Au MF. At the minimum laser power of 2.8 µW, a distinguishable current difference ($\Delta I \pm \sigma_{noise}$) of 1.08 pA ± 0.24 pA (RMS) was observed compared to the current trace under dark conditions. This suggests detection of the lowest reported current step signal of ~1 pA in bulk electrolyte media[84], which corresponds to an improved signal-to-noise ratio (SNR) of ~ 4.5 at a reciprocal response time ($\Delta t^{-1}$) of 1 s$^{-1}$ (0.5 Hz cut-off frequency) even in the absence of a Faraday cage.[47] This enhanced current resolution enables detection of subtle photo-induced chemical reactions corresponding to 16.5 µM concentration of Fe(CN)$_6^{3-}$ oxidant molecules, with a noise-limited detection threshold of around 3.67 µM. This represents a high sensitivity of 65.4 nA/M, and an improved limit of detection (LOD) of 11.0 µM, surpassing the performance of bulk electrochemical techniques that rely on highly-sensitive optical probes (sub-mM)[85] and micro-scale SECM-AFM Pt electrochemical probes (tens of µM)[86], and pushing the limits towards the best values reported for electrochemical methods in confined gap space (fM to pM)[87–90]. Furthermore, when normalized to electrode size, our nanoelectrodes exhibit an LOD$_{area-normalized}$ (75.8 µM/µm²) over 22 times larger than the average LOD$_{area-normalized}$ of 1µm-radius microelectrodes (ca. 3.5 µM/µm²) at almost the same LOD values,[6] indicating superior sensitivity in confined regions, making them well-suited for nanoscale electrochemical imaging. The authors attribute this improvement to the optimized nanoelectrode design, which includes a thin glass wall and single-crystalline structure, enhancing the collection of photo-generated molecules by improving mass transport rates and closer tip-to-substrate distances, and minimizing the fundamental electrochemical noises due to higher crystal quality and low-energy crystal defects (i.e. coherent {111} Σ3 twin boundaries).

Finally, and most remarkably, we demonstrate the extraordinary versatility of our chemical growth approach for fabricating not only Au NEs, but also Au UMEs, platinum NEs and multifunctional NEs tailored for advanced electrochemical microscopy techniques (see **Figure 4l-o**). For example, by simply substituting the gold precursor with chloroplatinic acid, we successfully fabricated Pt NEs under identical growth conditions (**Figure 4l**), showcasing the adaptability of this method. Beyond gold and platinum, this polyol-based growth strategy holds immense potential for other metals and alloys, including Ag[91], Cu[92], Pd[93], and even Bi[94]—materials critical for electrochemical research yet unattainable via traditional laser-pulling methods[95].

Furthermore, we demonstrate that performing a partial cross-sectional FIB cut on disc-shape Au NEs can expose a recessed disc-shape nanoelectrode and result in an L-shape tip geometry with a protruding insulating glass tip capable of simultaneously detecting atomic forces and Faradaic processes (**Figure 4m**). To the best of our knowledge, this represents the most efficient and high-throughput method to date for fabricating dual-functional tips for combined SECM-atomic force microscopy (SECM-AFM),[96,97] paving the way for advanced research that require high-resolution constant-distance electrochemical and tomographic imaging.

We additionally demonstrate complete chemical growth within theta glass nanopipettes having non-symmetric geometries, which is remarkable for fabrication of high performance thin-wall double barrel (**Figure 4n**) and dual (**Figure 4o**) metal NEs. These versatile electrodes enable simultaneous detection of Faradaic currents alongside ionic currents, supporting concurrent analysis of surface reactivity[98], product selectivity[99], surface topography[100], pH[101], temperature[102], surface charge[103,104], and chemical delivery[105,106] in combined SECM, scanning ion conductance microscopy (SICM), and scanning electrochemical cell microscopy (SECCM) techniques. Until now, such combined methods have largely been limited to carbon NEs[100,105,107] with limited electrochemical performance for inner-sphere reactions or relied on surface-modified carbon NEs[108,109] due the difficulties with fabrication of multifunctional metal NEs with the laser pulling approach[110].

Overall, the remarkable material and geometric versatility of our approach not only expands the scope of NE fabrication but also paves the way for the development of next-generation electrochemical microscopy and innovative applications across diverse fields by bridging between electrochemistry, biology, and spectroscopy.

**Conclusions**

These results represent a significant breakthrough in scalable fabrication of high-quality, thin-wall Au NEs, addressing longstanding challenges in reproducibility, performance reliability, and versatility. By combining optimized laser pulling and chemical growth techniques, together with a precise FIB cut process, we obtained exceptional reproducibility (~80%) in the fabrication of thin-wall Au NEs with controllable sub-100 nm radii core sizes and low RG values (~1.2 to ~5). The method also shows an extraordinary versatility for high-throughput fabrication of other metal/alloyed NEs (e.g. Pt, Ag, Cu, and even Bi), thin-wall UMEs, and multifunctional NEs including dual, double-barrel, and recessed L-shape NEs for various electrochemical probe microscopy techniques, from SECM to EC-STM, AFM, SICM, and SECCM hybrid methods. Comprehensive structural and electrochemical analyses on disc-shape Au NEs confirmed a gap-free gold-glass interface together with a unique single-crystalline nature of the Au core, resulting in a high electrochemical quality and stability in electrolyte environments. A size dependency was observed for the electrodes with a more positive standard potential, higher $C_{dl}$ value, and higher HER activity for smaller size electrodes, the latter being of high significance for improving the signal-to-noise ratio with small-size NEs. We demonstrated the practical potential of these NEs in photo-SECM imaging and detection of uniquely low current step signals (1.08 pA ± 0.24 pA) for high sensitivity monitoring of weak photo-oxidation reactions on single-crystalline Au micro-flake surfaces at outstanding detection sensitivities (65.4 nA/M) and limit of detections (11.0 μM), sub-200 nm spatial resolutions, and extremely stable performance for over 7 hours of photochemical imaging. By addressing long-lasting issues in the versatile fabrication of high-quality thin-wall metal NEs/UMEs, this work paves the way for the development of next-generation spectroelectrochemical and near-field microscopies towards high-resolution single-molecule and single-nanoparticle analysis. This also opens completely new possibilities in nanoelectrochemisty, material science and biochemistry by providing an excellent single-crystalline platform for operando fundamental research into interfacial phenomena, such as double-layer structure[53], nanoscale charge transport mechanisms[72], as well as single-enzyme kinetics[111,112], neurotransmissions[77], and biomolecular interactions[113].

## Methods

**Fabrication of Nanopipettes**
Single-barrel nanopipettes were fabricated from borosilicate glass capillaries (1.2 mm outer diameter, 0.69 mm inner diameter; Science Products GmbH) having no filament inside. The capillaries were cleaned by sonication in acetone and ethanol for 10 min, followed by several times rinsing with ultrapure water and drying in an oven at 100 ˚C before pulling. No hazardous Pirhana or silanization treatments were used. The nanopipettes were fabricated by using a combined laser shrinking and laser pulling method on a laser-based P-2000 pipette puller (Sutter Instruments). First, the glass capillaries were locally shrank at the center using high energy of the $CO_2$ laser for increasing the wall thickness and thus the strength of the capillary before the pulling process. In this step, a pair of home-made stoppers were used to fix the pulling arms and then to controllably soften and shrink the glass capillary. Next, after removing the stoppers, an optimized two-line program was used to hard pull the pre-shrank capillaries into ultrafine nanopipettes. At constant Heat parameters, the duration of the shrinking Time and the Delay parameter in the hard pulling step were separately used to control the achievable orifice size and the shank/taper length of the nanopipettes, respectively. Sub-20 nm-radius short-taper nanopipettes were obtained at a > 90% repeatability by a 90 s shrinking time and a 190 Delay parameter in the hard pull step. The orifice size of the fabricated nanopipettes was characterized under SEM without using any conductive coating. Image integration at low electron beam energy, i.e. 2 kV, was used for SEM imaging with the least charging effect without using any conductive coating material.

**Fabrication of Gold Nanoelectrodes**
Gold nanoelectrodes were fabricated by a polyol-based chemical growth process[54] inside ideal-shape glass nanopipette templates (20 nm-radius and 250 μm-length taper). First, the tip of the glass nanopipette was dipped in pure EG, and then the glass nanopipette was backfilled with a growth solution containing 0.125 M $HAuCl_4$ in EG. This resulted in an advantageous separation, i.e. an air bubble, between the backfilled growth solution and the tiny front-filled EG. The filled nanopipettes were then vertically hold in 6 mL glass vials containing a bulk solution of 2 mL EG and 0.5 mL 200mM $NaBH_4$ in ethanol. Next, the vials were immediately transferred to a muffle furnace (Nabertherm GmbH) heated at 110 °C for starting chemical growth for 24h in all nanopipettes in a batch. This condition enabled selective growth of continuous ~300 μm-length Au deposits at an > 80% repeatability with the least amount of unwanted particles and chemical etching on both exterior and interior surfaces. The NEs were then naturally cooled down in the furnace and rinsed by ultrapure water and ethanol for removing the bulk solution residues from their surface. A cleaning treatment with ultrapure water and ethanol was then performed on the as-grown NEs in order to remove the remained growth solution and any randomly formed unwanted particles, if any, at the nanopipette neck, behind the solution/air interface of the air bubble. Next, a long-taper W micro-contact was physically connected to the Au deposit under an optical microscope to establish a soft electrical contact. The Au NEs were finally perfected by complete FIB cutting of the protruded Au or etched glass imperfect parts into disc-shape electrodes. A 30 kV and 2 pA Ga ion beam was used for sharp cutting of the NEs.

**Fabrication of Long-taper W Micro-contacts**
Long-taper W micro-contacts were prepared by using a very facile two-step electrochemical approach consisting of thinning and cutting steps. A W microwire (25 μm diameter, Goodfellow, 99.9%) was first vertically immersed by 3.5 mm into a 10 mL of 2 M KOH etchant solution using a manual microstage. In the first step, a 0.5 V DC potential was applied between the W wire and a Cu wire loop cathode 1cm below the solution surface. This results in oxidative dissolution of $WO_2^-$ anions in water, especially slightly below the meniscus at the air-electrolyte interface, leading to a necking shape and a long uniformly thinned section that can finally drop off at the neck in a complete etching process.[114] The etching process was controlled by monitoring the current level until a ~50% current drop was observed, and stopped by stopping the bias. At this step, the W micro-wire is incompletely etched and thinned down to an ~5 μm diameter. Then, the etched W wire was lifted up by ~1 mm and biased at a lower potential, e.g. 0.2 V. An ~700 μm-length W wire was cut and tapered at this step after the drop off of the bottom part controlled by a sudden current drop. The fabricated W micro-contact was finally rinsed with DI water to terminate the etching reaction. In this method, the current drop criteria at the first step and the amount of the upward lift at the second step separately allow a good control over the width and the taper length of the W micro-contact, respectively. An > 90% reproducibility was achieved for fabrication of ideal-profile (5 μm diameter 700 μm taper length) W micro-contacts for Au NEs.

**Electrochemical Testing of Gold Nanoelectrodes**
All voltametric experiments were performed in a one-compartment three-electrode cell connected to a CHI 760E potentiostat (CH Instruments) without any Faraday cage. A leak-free Ag/AgCl reference electrode and an Au counter electrode were used to avoid cross-contamination of the Au NEs with $Ag^+/Cl^-$ ions and/or Pt during electrochemical testing. To remove oxygen, the electrolyte solutions were purged with $N_2$ gas for 30 min before the experiments. First, outer-sphere voltammetry was performed in a 2 mM $Fe(CN)_6^{4-}$ in 0.25 M $Na_2SO_4$ solution for evaluation of the electrical connection and geometrical surface area. The effective radius of the electrodes was evaluated from the measured diffusion-limiting currents through the modified Saito's equation[63,64] (see eq.1) for different RG values estimated from SEM imaging. Next, the electrodes were activated by multiple cycling in 50 mM $H_2SO_4$ solution within a 0 to 1.4 V vs Ag/AgCl potential window at 100 mV.s$^{-1}$ scan rate for about 30 cycles until a pristine surface was produced. Cycling was stopped once a stable CV response was observed. Steady-state voltammetry of the electrodes was measured at a 10 mV.s$^{-1}$ scan rate within 0.2 to 1.4 V and -0.8 to 0.8 V vs Ag/AgCl potential windows for evaluation of the Au redox processes and HER activity, respectively. The electrochemical surface area and roughness factor (RF) of the electrodes were calculated from the characteristic values of the consumed charge for the reduction of an AuO monolayer in a one-to-one ratio (390 μC.cm$^{-2}$) and the ratio of the calculated microscopic surface area to the geometrical area ($A_m/A_g$), respectively.[61] The double-layer charging current, $i_c$, was background subtracted for the surface area calculations due to the low current level of the Au redox process on NEs. The sealing quality of the electrodes was electrochemically characterized by performing CVs in the double layer and HER regimes and evaluation of the CV profiles.[69] Double layer capacitance ($C_{dl}$) values in F cm$^{-2}$ were obtained by calculating the $i_c$ at 0.3 V vs Ag/AgCl from double layer CVs and by using $C_{dl} = |\Delta i_c|/2A_m.v$, where $A_m$ is the microscopic area of the electrode in cm$^2$, and v is the scan rate in V s$^{-1}$. Unless otherwise specified, all the experiments were repeated three times, and only the second cycles were reported for ensuring a steady-state response condition. A fast Fourier transform (FFT) filter was used for smoothing the high-scan rate measurement data.

Singal-to-noise ratio (SNR) for nanoelectrodes was calculated by dividing the light-induced change in tip current, i.e. current step amplitude ($\Delta I$), to the corresponding standard deviation of the measured background current noise, i.e. noise level ($\sigma_{noise}$), based on the measurement reported in **Figure 4k**. To reduce high frequency environmental noise, a 0.5 Hz low pass filtering was applied to the measurement ($\Delta t = 1/2f = 1$s). Detection sensitivity ($S$) and limit of detection (LOD) were obtained by using the modified Sato's equation (eq. 1) for $C^*$ calculation, and the following equations,[86]

$$S = \frac{\Delta I}{C^*} \quad \text{(eq. 3)}$$

$$LOD = \frac{3\sigma_{noise}}{S} \quad \text{(eq. 4)}$$

where $S$ is determined from the slope of the $\Delta I$ vs. $C^*$ curve, derived from data measured at different laser powers.

**Synthesis of Gold Micro-flakes**
High-quality single-crystalline gold micro-flakes were directly grown on borosilicate glass substrates by a halide and gap-assisted polyol process.[54] A PMMA wet-transferring method[115] was used for transferring the Au MFs onto a $TiO_2$/ITO-coated glass substrate. The sample was then exposed to an oxygen plasma (2 minutes, 500 W; Tepla 300) to remove any PMMA residue left from the transferring step.

**Transmission Electron Microscopy**
TEM analysis was performed using a FEI Tecnai Osiris microscope operated at 200 kV. BF-TEM images were recorded on a Gatan Orius camera. Crystal orientation mappings were performed using an Astar system[59]. Specifically, an electron probe of 14 pA was precessed on the sample at an angle of 1° at 100 Hz and raster-scanned with a 2.5 nm step size and 50 ms dwell time. Local precession-assisted diffraction patterns were collected with a Stingray camera recording the phosphor screen at a camera length of 165 mm. The Astar v2.2 software suite was used to index the pseudo-kinematic diffraction patterns of the Au face-centered cubic material. Finally, the data representation and analysis used MTEX v5.10.2 toolbox for MATLAB[116].

**Scanning Photo-electrochemical Microscopy**
Photo-SECM imaging was performed using a home-made SECM instrument on an inverted optical microscope for back illumination of the sample. Measurements were done in a three-electrode cell containing a leak-free Ag/AgCl reference electrode and a Pt counter electrode in a 4 mM $Fe(CN)_6^{4-}$ in 0.25 M $Na_2SO_4$ electrolyte solution. Tip Z scanning and sample XY scanning were realized by using Nano-OP and Nano-LP XY piezo-stages (Mad City Labs). Electrochemical measurements were done by a VSP300 potentiostat (Biologic). The Au NE tip was biased at a diffusion-limiting 0.4 V vs Ag/AgCl potential, and positioned at a distance corresponding to a 25% setpoint value in a negative feedback approach curve. An approach curve fitting was used by employing analytical approximations reported in [117]. Constant-height SECM imaging was performed by illumination of the sample with a 516 nm focused laser (10 µW) and applying an oxidative tip potential, i.e. 0.4 V vs Ag/AgCl, which satisfies a competition mode of operation for a photo-oxidation reaction on the substrate. The beam size was about 1 µm, and the sample was scanned at 200 nm step sizes. All the data were recorded using in-house programs written in LabView (National Instruments), and were plotted with Origin Pro software with no extra smoothing method.

**Acknowledgments**

M.S. and G.T acknowledge the support of the STI Imaging Fund, supported by the EPFL Center for Imaging. M.S, F.K. and G.T acknowledge the support of the Swiss National Science Foundation (Eccellenza Grant #194181). The authors also acknowledge the Laboratory for Bio- and Nano-Instrumentation (LBNI) and Laboratory of Biomaterials for Immunoengineering (LBI) for providing access to a Laser puller and centrifuge machines, respectively. Finally, the authors would like to thank Dr. Marcos Penedo Garcia for his advice on FIB, Dr. Alan Bowman for his advice on optical microscopy of the electrodes, Dr. Kiseok Oh for his advice on the performance of the electrodes, Dr. Priscila Vensaus for her comments on the manuscript, and Prof. Georg Fantner, and Mr. Barney Frederick Drake, and Mr. Paul Feurstein for their help with SECM setup development.